\documentclass[%
 aip,
 amsmath,amssymb,
 reprint,%
]{revtex4-1}
\usepackage{booktabs} 
\usepackage{graphicx}
\usepackage{dcolumn}
\usepackage{bm}

\usepackage[utf8]{inputenc}
\usepackage[T1]{fontenc}
\usepackage{mathptmx}
\usepackage{etoolbox}
\usepackage{xcolor}

\makeatletter
\def\@email#1#2{%
 \endgroup
 \patchcmd{\titleblock@produce}
  {\frontmatter@RRAPformat}
  {\frontmatter@RRAPformat{\produce@RRAP{*#1\href{mailto:#2}{#2}}}\frontmatter@RRAPformat}
  {}{}
}%
\makeatother
\begin{document}

\preprint{AIP/123-QED}

\title[Sample title]{Frequency-Dependent Magnetic modulation of deposition morphology}

\author{Sunil Kumar Saroj}
 \altaffiliation[Also at ]{Departmrnt of mechanical engineering, IIT Kanpur}
\author{Pradipta Kumar Panigrahi}%
 \email{sunilkrsiitk@gmail.com}
\affiliation{ 
Departmrnt of mechanical engineering, IIT Kanpur
}%

\date{\today}

\begin{abstract}

This paper presents a novel approach for magnetic modulation of deposition morphology in an evaporating ferrofluid droplet. The magnetic field strength and ferrofluid concentration are kept unchanged, while the actuation frequencies are varied from 0.016 Hz to 5 Hz. In the absence of a magnetic field, a coffee-ring formation is observed and consistent with previous studies\cite{deegan1997capillary,deegan2000contact,saroj2019drying}. The application of a time-dependent magnetic field significantly modifies the deposition morphology.  The periodic magnetic field induces the formation of multiple concentric rings during evaporation. The number of rings initially increases with increasing actuation frequency of the electromagnet. However, beyond a critical actuation frequency ($f_c = 0.2\,\text{Hz}$), the number of rings decreases. At higher actuation frequencies, magnetic particles preferentially deposit in the central region of the droplet, resulting in suppression of the coffee-ring effect. Additionally, the thickness of the inner rings and the ring spacing decrease with increasing actuation frequency up to critical actuation frequency. The transition from multi-ring formation to coffee-ring suppression is governed by the competition among magnetic forcing, capillary flow, and particle diffusion. The underlying physical mechanisms responsible for droplet dynamics and deposition morphology under periodic magnetic fields are evaluated using scaling arguments. The results demonstrate that diffusive particle transport plays a dominant role in determining the deposition pattern. A non-dimensional magnetic switching number, based on the magnetic perturbation timescale, is introduced as a control parameter to characterize the frequency-dependent deposition behavior.

\end{abstract}

\maketitle

\begin{quotation}
\end{quotation}

\section{\label{sec:level1}Introduction}

The ordered assembly of drying droplets containing micro- and nanoparticles on solid surfaces has numerous important applications, including inkjet printing\cite{tiberto2013magnetic}, bioassay manufacturing, painting, spin coating, and various chemical and biological processes \cite{kumar2017fabrication}. In particular, the ordered assembly of magnetic nanoparticles offers promising applications in magnetic recording technologies, micro--sensor devices\cite{men2015controlled}, biosensing and diagnostics, memory bit expansion, fabrication of micro--electrodes, magnetic tracks and templates\cite{yellen2005arranging}, water purification \cite{pinto2020application}, and the cores of radio-frequency identification (RFID) resonators. A key advantage of magnetic nanoparticle deposition is that the particles can behave like non-magnetic particles in the absence of a magnetic field, while exhibiting a strong and controllable response when an external magnetic field is applied. Such external actuation of magnetic nanoparticles enables low-cost, contact-free manipulation, requires no moving mechanical parts, and provides precise control over particle transport and final deposition patterns.

Magnetic nanoparticles aligned along the direction of an applied magnetic field can form chain-like structures or interconnected networks. These assemblies exhibit a tunable response depending on the magnitude and direction of the applied field \cite{yang2023recent}. Such anisotropic characteristics facilitate direction-dependent electrical, thermal, and mechanical properties. Consequently, magnetic nanoparticles can be effectively utilized in printed sensors, strain-sensitive devices, flexible circuits, and tunable photonic materials \cite{erb2013self,maier2007plasmonics,bhattacharjee2026evaporative}.

\par
The coffee-ring–like deposition pattern is typically formed during the evaporation of a sessile droplet with a pinned contact line. This phenomenon arises from the divergence of the evaporative flux near the contact line, which induces an outward capillary flow that transports suspended particles toward the droplet periphery, resulting in particle accumulation and deposition at the edge region of the droplet \nocite{deegan1997capillary,deegan2000contact,marc2022coffee}. Consequently, a key challenge in evaporative deposition processes is the suppression of the coffee-ring effect under pinned contact line conditions. The motion of particles inside a drying droplet governs the final shape and size of the deposition morphology. Therefore, particle motion can be effectively controlled by regulating the evaporation process through substrate modification, substrate heating, variation of solvent properties \cite{seifert2015additive,sliz2020taming} or application of the external agency on the particles such as magnetic field\cite{bashtovoi1987magnetic,dormann1997magnetic,saroj2020magnetic}, acoustics \cite{mampallil2015acoustic}, adding air bubbles in droplets\cite{yang2020air} and electrowetting\cite{eral2011suppressing}. Lim et al. \cite{lim2009experimental} investigated the spreading dynamics and evaporation behavior of on-demand jetted droplets generated using a piezo-driven method on substrates maintained at different temperatures. They observed that, for water droplets, spreading was completed before the evaporation process ended, whereas for ethylene glycol droplets, spreading and evaporation overlapped at higher substrate temperatures due to their higher wetting characteristics. Furthermore, it has been demonstrated that the deposition pattern of a drying droplet can be transformed from a coffee-ring structure to a more uniform deposit by increasing the substrate temperature, owing to the modification of internal flow and evaporation dynamics \cite{he2017controlling}.
\par
Kolchanov et al.\cite{kolchanov2019sol} have recently investigated the inkjet printing of magnetic material on substrates of different wetting properties. The magnetic properties of the deposit after drying have been measured using magnetic force microscopy, which demonstrated a good magnetic response for use in anti-counterfeiting and magnetic pattern printing. They did not use a magnetic field during the inkjet printing and drying processes. The application of a magnetic field during the drying droplet containing a magnetic field significantly reduces the coffee ring effect with the formation of a chain of magnetic nanoparticles \cite{al2019inkjet,saroj2020magnetic}. The particles get aligned in the direction of applied magnetic field which may increase the high frequency permeability and decrease the hysterisis loss \cite{song2014inkjet}. 

Wu et al. \cite{wu2018multi} proposed a theoretical model to explain multi-ring formation in drying droplets, demonstrating that the emergence of multiple rings depends on the evaporation rate, equilibrium contact angle, and receding contact angle. The stick–slip motion of the contact line during the evaporation process is identified as the primary mechanism responsible for multi-ring generation \cite{moffat2009effect}. Weon and Je \cite{weon2010capillary} reported that multi-ring patterns can also arise from differential particle deposition during droplet drying, which is strongly influenced by particle size. Furthermore, Marc et al. \cite{marc2022coffee} demonstrated multi-ring formation in droplets containing magnetic nanoparticles coated with (3-aminopropyl)triethoxysilane (APTES) evaporating on a low-density polyethylene (LDPE) film under an externally applied magnetic field. In their study, multi-ring patterns emerged due to the stick–slip motion of the contact line during droplet evaporation. An enhanced concentration of particle deposition at the ring locations was observed when a magnet was placed beneath the droplet, with the deposition intensity depending on the magnet–droplet separation distance. Notably, in this approach, the droplet was not pinned during evaporation, and multi-ring formation could also occur even in the absence of a magnetic field. Therefore, it is essential to develop a technique that enables controlled multi-ring formation while suppressing the coffee-ring effect under pinned contact-line evaporation. The magnetic actuation timescale may play a crucial role in regulating contact line dynamics and particle transport, thereby enabling controlled assembly of colloidal magnetic nanoparticles during droplet evaporation.

The present study reports the manipulation of magnetic particles within an evaporating sessile droplet using a periodically applied magnetic field. A pointed electromagnet is positioned above the droplet apex. The separation distance between the droplet apex and the electromagnet tip, as well as the applied current, are maintained constant throughout the experiments. The periodic application of the magnetic field significantly modifies the motion of magnetic particles within the droplet, resulting in the attenuation of particle deposition at the droplet periphery and the formation of multiple concentric rings inside the droplet. Experiments are conducted over a range of electromagnet switching frequencies. Multi-ring deposition patterns are observed, accompanied by a strong suppression of the peripheral coffee-ring deposit with increasing actuation frequency. A critical switching frequency, $f_c = 0.2$ Hz, is identified, at which the maximum number of concentric rings is formed. The formation of these multi-ring patterns is attributed to the continuous contraction and spreading of the droplet during the final stage of evaporation, coupled with frequency-dependent diffusive transport of particles. This interplay leads to the successive deposition of particles in the form of multiple concentric rings. A magnetic switching number, defined based on the diffusion time scale and the forcing frequency, is introduced to quantify the different deposition morphologies. The physical mechanisms governing the observed deposition behavior are explained using scaling arguments.

\begin{figure}
\includegraphics[width=0.5\textwidth]{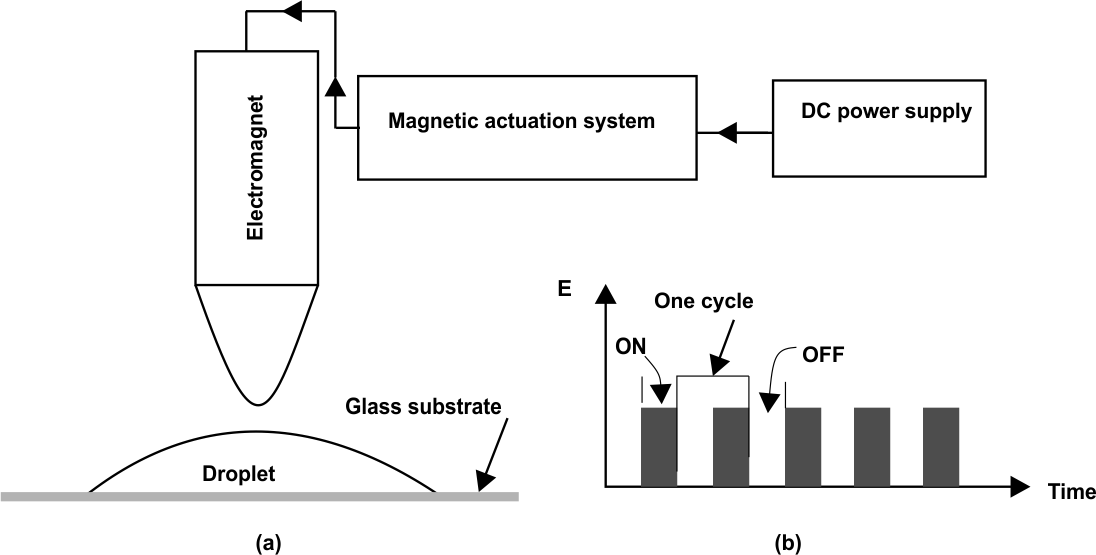}
\caption{\label{fig:setup} Sketch of (a) experimental arrangement and (b) shows the ON and OFF of electromagnet with time.}
\end{figure}

\subsection{\label{sec:level2}Experimental setup}

Figure \ref{fig:setup} (a) shows the experimental arrangement for the droplet-electromagnet arrangement. A ferrofluid droplet of 1.0 $\%$ concentration placed on the hydrophobic glass surface. The volume of the droplet is equal to 2.0 $\mu$l. A pointed electromagnet is placed over the droplet at a distance of 0.35 mm from the apex of the droplet. The current is supplied to the electromagnet by the DC power supply to generate the magnetic field. An electrical circuit is used to control the actuation frequencies of the electromagnet. The evaporating side view visualization of the droplet is done by the goniometric arrangement. The images are captured by the goniometric arrangement at 10 s interval time. The detailed information about the goniometric arrangement can be found in our earlier article \cite{saroj2020magnetic}.  The top-view images are captured after complete evaporation using the Confocal microscopy system. 

Figure \ref{fig:setup} (b) shows the switch ON and switch OFF of the electromagnet during course of the evaporation of the droplet. The switch ON and switch OFF periods are controlled by the Arduino programming. The time depended square magnetic field has been utilized to control the deposition morphology for this study and can be described as following;

\[
B(t)=
\begin{cases}
B, & 0 \le t < \frac{T}{2} \rightarrow \text{switch ON} \\
0,   & \frac{T}{2} \le t < T \rightarrow \text{switch OFF}
\end{cases}
\]
Here, $T$ denotes the magnetic actuation period, while the corresponding actuation frequency is given by, $f = \frac{1}{T}$

The contact angle of the water droplet on the glass substrate which is being used in the present study is 94$^{\circ}$ \cite{saroj2021magnetophoretic}.

\section{Results and discussion}

The primary objective of the present study is to experimentally demonstrate the generation of multiring patterns as well as the suppression of the coffee-ring effect in an evaporating ferrofluid droplet on a glass substrate. The observations of the current study are organized as follows: (1) deformation and shape evolution of the magnetically actuated ferrofluid droplet, (2) particle deposition morphology within the droplet, and (3) discussion of the underlying physical mechanisms.

\subsection{\label{sec:level3}Dynamic Shape Evolution}

The magnetic particles inside the droplet behave like non-magnetic particles during evaporation in the absence of an applied magnetic field. Upon switching ON magnet, the particles become magnetized and align along the direction of the applied magnetic field. Consequently, the particles migrate toward regions of higher magnetic field strength, primarily toward the droplet apex, where the magnetic field is maximum. Simultaneously, the droplet undergoes deformation, resulting in changes to the contact angle and height. For the current study, the magnetic field strength is carefully selected to avoid droplet splitting. While splitting was observed at higher particle concentrations, those cases are not discussed here since they are not relevant to the primary objectives of this work.

\begin{figure*}
\includegraphics[width=1.0\textwidth]{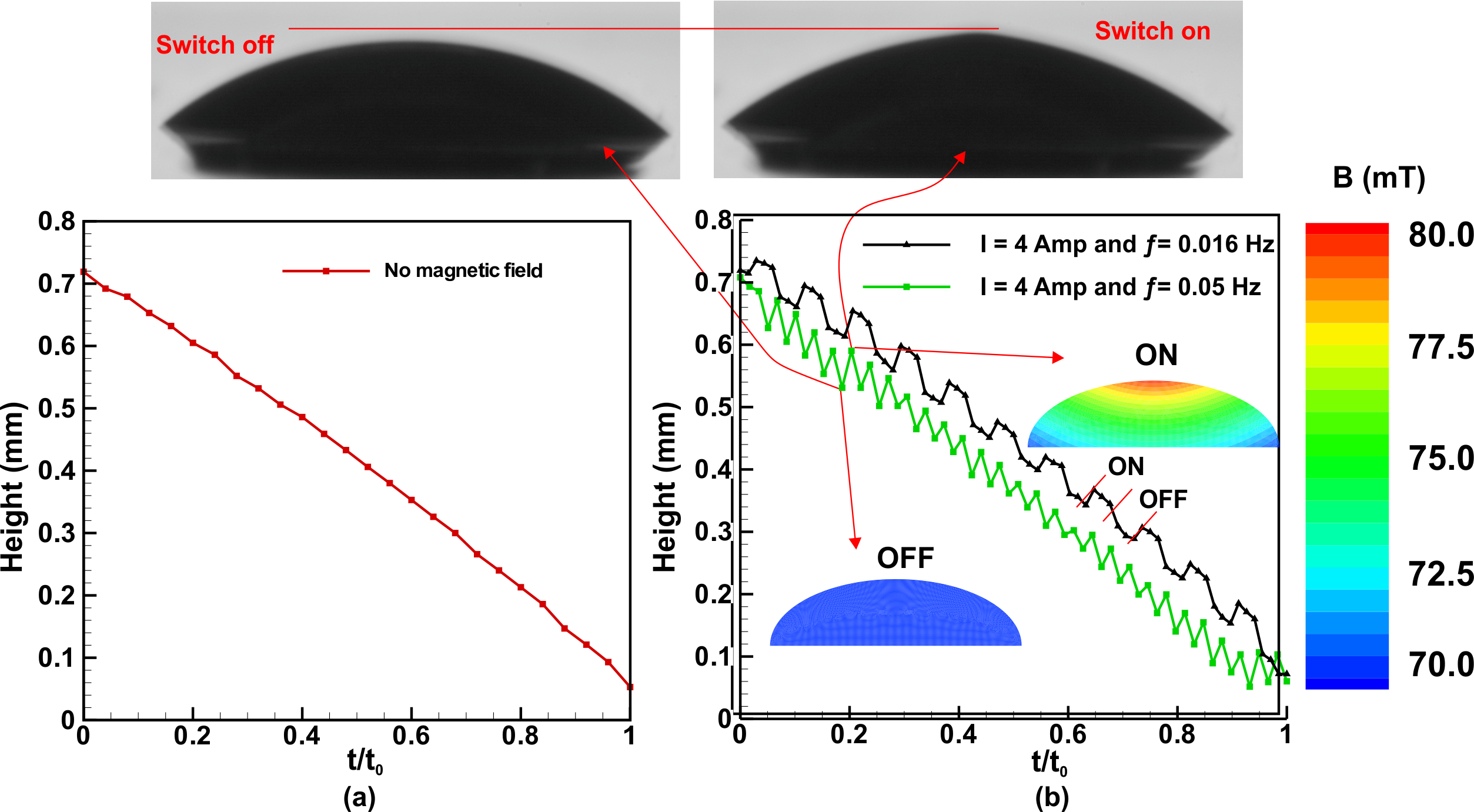}
\caption{\label{fig:Height} Variation in height during evaporation of the droplet: (a) without application of the magnetic field and (b) with actuation of the magnetic field at $f$ = 0.016 Hz and $f$ = 0.05 Hz. The side view images (top of the figure) show the increased height during switch on state and switch off states of the magnet during magnetic actuation. The inset color images in figure (b) show the magnetic flux density (B) distribution inside the droplet by COMSOL Multiphysics.}
\end{figure*}

Figure~\ref{fig:Height} shows the experimentally measured variation in droplet height during evaporation, both without and with magnetic field actuation.  Figure \ref{fig:Height} (a) shows the height decreases linearly over the time without any actuation of the magnetic field during course of evaporation. This variation of height shows consistency with the previous studies.

Figure ~\ref{fig:Height}(b) shows the temporal variation of the droplet height during the switching ON and OFF of the electromagnet over the course of evaporation. The corresponding side-view images, shown at the top of Fig.~\ref{fig:Height}, illustrate the droplet shape at a switching frequency of 0.05~Hz for the electromagnet switched OFF and ON, respectively. The initial contact angle is $\theta = 49^\circ$ in the switched-OFF state, while it reduces to $\theta = 45^\circ$ during the switched-ON state. The inset color images in  Fig.~\ref{fig:Height} (b) show the magnetic flux density distribution, $B = \mu_0 (M + H)$, inside the droplet, obtained using COMSOL Multiphysics. When the magnet is switched off, the droplet (shown in blue) experiences no magnetic field. Upon switching on the magnet, the magnetic field is strongest near the apex and decreases towards the contact line. The simulations were carried out following the methodology described in our article \cite{saroj2016two,saroj2020magnetic}, to which the reader is referred for further details.

The droplet height increases during the switch ON state of the electromagnet and abruptly returns to its initial value when the field is switched OFF. This cyclic behavior persists throughout the evaporation process, as evident in Fig.~\ref{fig:Height}(b). With increasing actuation frequency, the number of such cycles increases, leading to larger and more frequent height fluctuations. The slight decrease in height observed at $f = 0.016$~Hz during the switch ON period arises from evaporation occurring over this relatively long duration. In contrast, when the frequency is increased to $f = 0.05$~Hz, no such decrease in height is observed during the switch ON period due to the shorter duration of the applied magnetic field. As a result, there is no significant reduction in droplet height during this interval. The height fluctuation is slightly larger for $t/t_0 \lesssim 0.45$ and for $t/t_0 \gtrsim 0.8$, irrespective of the switching frequency, while smaller fluctuations are observed for $0.45 \lesssim t/t_0 \lesssim 0.8$. This behaviour arises from variations in the magnetic field strength caused by changes in the relative distance between the magnet tip and the droplet apex, as well as by the increase in the local ferrofluid concentration during evaporation. The magnetic field strength is higher for smaller distance between the magnet and the droplet apex and as the local particle concentration increases during final stage of evaporation, leading to increase in magnetic field strength which causes larger height fluctuations for $t/t_0 \lesssim 0.45$ and $t/t_0 \gtrsim 0.8$. The authors refer the reader to our earlier article~\cite{saroj2020magnetic} for further details.

\begin{figure}[htb!]
\centering
\includegraphics[width=0.5\textwidth]{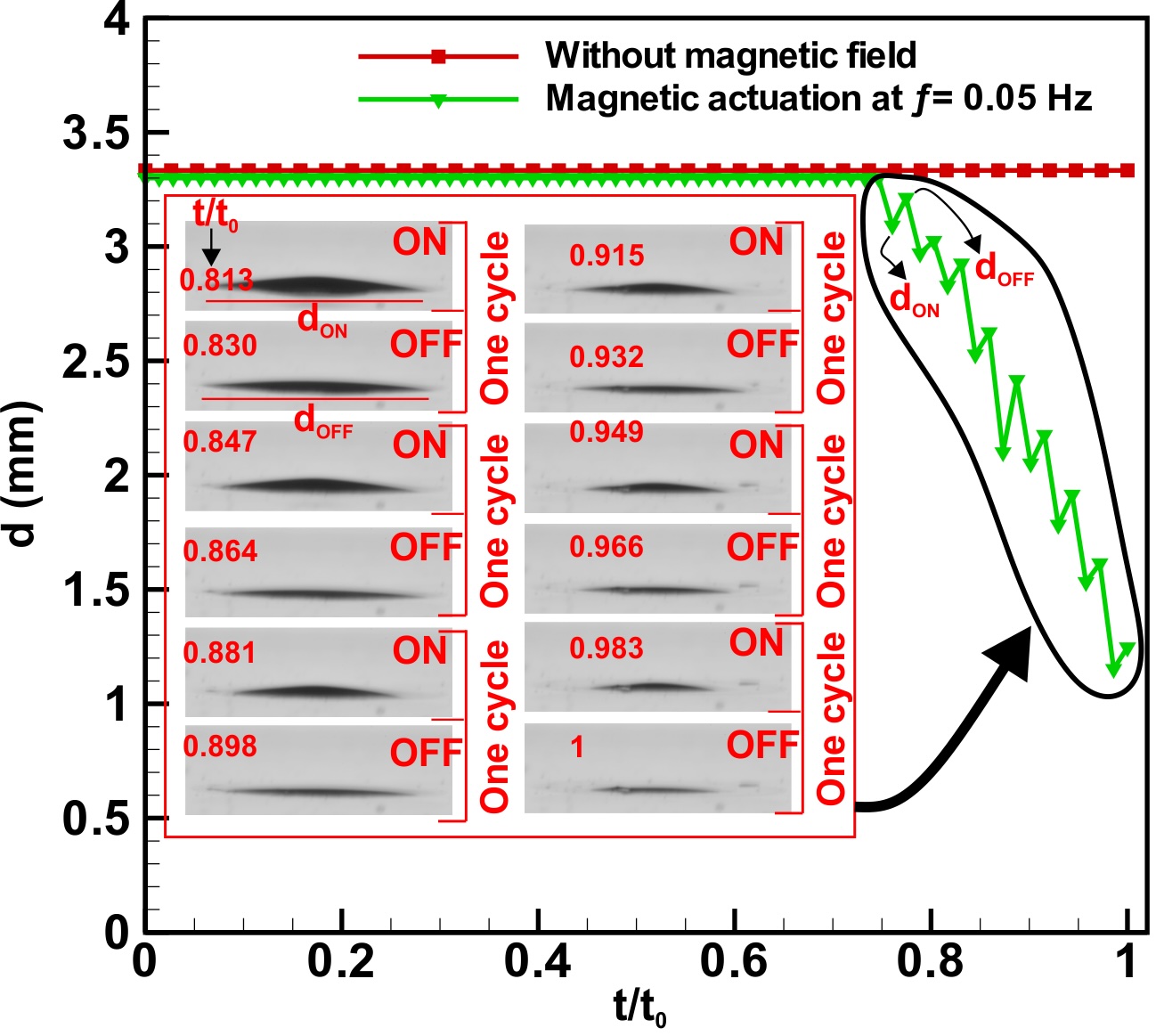}
\caption{\label{fig:dia} Variation of the diameter of the droplet with evaporation time for without magnetic field actuation and with magnetic field actuation. .Here, $\rm{d_{ON}}$ and $\rm{d_{OFF}}$ are diameters of the droplet during ON and OFF of electromagnet, respectively.}
\end{figure}

Figure \ref{fig:dia} shows the variation of the contact line diameter of the droplet with evaporation time. The contact line is pinned during all evaporation time in the absence of magnetic field actuation. With magnetic actuation, the diameter of the droplet changes after approx 80 $\%$ evaporation time (during rush hour evaporation). During this period, the diameter reduces to the smaller diameter ($\rm{d_{ON}}$) and increases to higher diameter ($\rm{d_{OFF}}$) for the ON and OFF periods in one cycle, which is clearly indicated in Fig.~\ref{fig:dia}. The alternate pinning and depinning of the contact line during the later stages of evaporation may contribute to the formation of multiple concentric rings, depending on the actuation frequency. The mechanism underlying this phenomenon will be discussed in a later section.

\subsection{Deposition morphology}

The evaporation of droplets containing micro- or nanoparticles on solid substrates under pinned contact line conditions leads to the formation of a characteristic coffee-ring pattern due to particle accumulation near the contact line. This phenomenon is driven by a radially outward capillary flow induced by non-uniform evaporation, wherein the evaporative flux increases from the droplet center toward the contact line. For small contact angles ($\theta \ll 90^\circ$), the local evaporative flux scales as \cite{deegan1997capillary,deegan2000contact}
\begin{equation}
J(r) \sim \frac{1}{\sqrt{R-r}} \quad \text{as } r \rightarrow R,
\end{equation}
which diverges near the contact line, resulting in enhanced particle deposition at the droplet periphery. 

\begin{figure*}[htb!]u5
\includegraphics[width=1\textwidth]{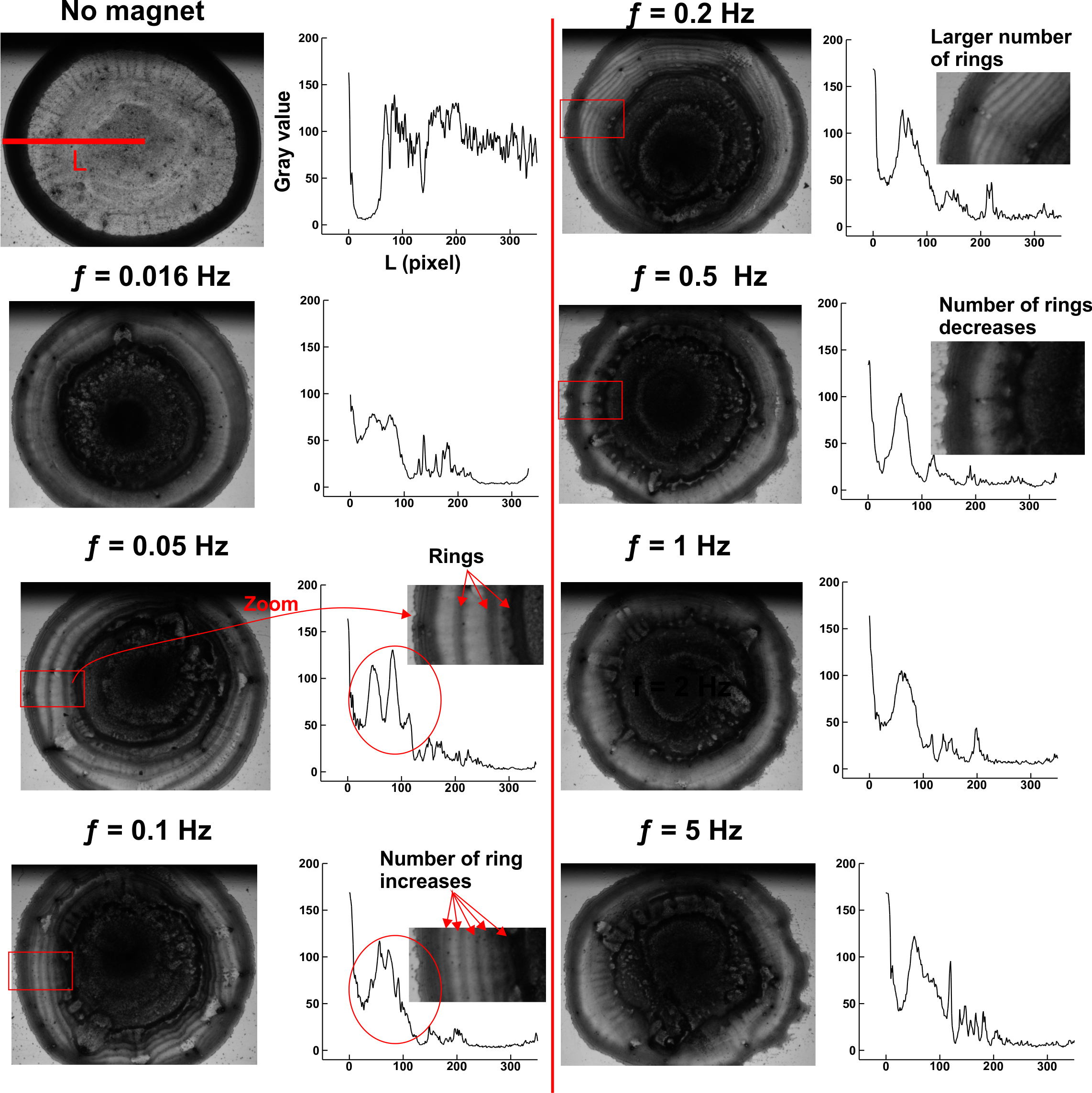}
\caption{\label{fig:1} Deposition pattern along with the variation in gray value of drying droplet at different switching frequencies.}
\end{figure*}

Figure~\ref{fig:1} shows the final deposition morphology of magnetic particles under different actuation frequencies, along with the case of no magnetic field actuation. In the absence of a magnetic field, particles assemble predominantly at the droplet edge after complete evaporation, forming a well-defined coffee-ring pattern consistent with previous reports (see Fig.~\ref{fig:1}, no magnet) \cite{deegan1997capillary,deegan2000contact}. To quantitatively compare particle distributions, the radial variation of gray value along the marked red line (L) in Fig.~\ref{fig:1} is presented for each case. All images were captured using identical confocal microscope acquisition settings to ensure a reliable comparison. A higher gray value corresponds to a lower particle concentration, whereas a lower gray value indicates a higher particle deposition density. For the no-magnet case, the gray value is minimum at the ring location and increases toward the droplet center, confirming strong edge-localized deposition.

At an actuation frequency of $f = 0.016~\mathrm{Hz}$, the particle distribution within the droplet is significantly altered, exhibiting enhanced deposition in the central region and a substantial reduction near the contact line. Both the deposition images and the corresponding gray-value profiles show a pronounced suppression of particle accumulation at the droplet edge. During the switched-ON period, particles migrate away from the contact line toward the droplet apex and do not return to the edge during the switched-OFF period. Only particles initially located very close to the contact line deposit at the periphery. In addition to the weak outer ring, a secondary ring of lower concentration is observed between the contact line deposit and a highly concentrated inner ring located at approximately $L \approx 115$ pixels. The most intense particle deposition occurs at the droplet center, as indicated by the sharp decrease in gray value for $L > 200$ pixels (Fig.~\ref{fig:1}, $f = 0.016~\mathrm{Hz}$). In total, three rings (including the outer ring) are observed.

When the actuation frequency is increased to $f = 0.05~\mathrm{Hz}$, a higher number of rings is observed, while the overall deposition behavior near the contact line and the droplet center remains similar to that at $f = 0.016~\mathrm{Hz}$ (see gray-value profile for $f = 0.05~\mathrm{Hz}$). The increased number of concentrated rings is attributed to the reduced duration of the switched-ON state of the electromagnet. As a result, particles near the contact line do not migrate sufficiently toward the apex and subsequently deposit during the switched-OFF period, leading to the formation of multiple highly concentrated rings. These rings predominantly form during the final stage of evaporation, when the contact line contracts during the switched-ON period and spreads during the switched-OFF period. A detailed discussion of this mechanism is provided in the following section. For $f = 0.05~\mathrm{Hz}$, four distinct rings are observed. 
With a further increase in frequency to $f = 0.1~\mathrm{Hz}$, the number of rings increases to six, accompanied by a reduction in the spacing between adjacent rings. This increasing trend continues up to $f = 0.2~\mathrm{Hz}$, where a maximum of 10 concentric rings are observed. This frequency corresponds to a critical switching frequency at which the number of rings is highest. Beyond this value, a further increase in $f$ leads to a reduction in the number of rings to two, which remains unchanged at higher actuation frequencies. The measured thickness of the concentring ring is equal to 42, 61, 51 and 28 $\mu$m corresponding to f = 0.016, 0.05, 0.1, 0.2, respectively.

Overall, the application of a periodic magnetic field at the apex of the droplet suppresses the conventional coffee-ring effect and promotes the formation of multi-ring deposition patterns. The number, spacing, and location of these rings are strongly governed by the magnetic field switching frequency.

\subsection{Discussion}

Mechanisms underlying the above reported experimental observations are described in this section

\subsubsection{ Mechanism of Droplet Shape Evolution}
The deformation of the droplet shape arises from the applied magnetic field and is opposed by restoring forces associated with surface tension and gravity. The relative importance of gravity and capillarity is characterized by the capillary length,

\begin{equation}
l_c = \sqrt{\frac{\sigma}{\rho_f g}} \sim 1.4 ~ \text{mm}
\end{equation}
Here, $\rho_f$, and $\sigma$ are 1043 $\frac{kg}{m^3}$ and 20 $\frac{mN}{m}$, respectively.
The corresponding gravitational Bond number, $Bo_n = \left(\frac{R}{l_c}\right)^2 \approx 1.33$,
indicating that surface tension and gravity are comparable. Consequently, the droplet retains an approximately spherical shape with a flattened sessile geometry.

Under magnetic actuation, the field pulls the droplet upward, while surface tension and gravity resist deformation. Surface tension opposes interfacial curvature changes, whereas gravity resists increases in droplet height. The capillary pressure resisting deformation, $P_{\sigma} = \frac{2\sigma}{R}$ and the hydrostatic pressure contribution scales as $P_g \sim \rho_f g h$. The magnetic body force acting on the droplet is

\begin{equation}
F_m = \mu_0 V_d (\mathbf{M} \cdot \nabla)\mathbf{H}.
\end{equation}
Using the scaling relations, $V_d \sim R^2 h, \quad \mathbf{M} \sim \chi \mathbf{H}$, the magnetic force can be approximated as $F_m \sim \frac{R^3 \mu_0 \chi H^2}{2h}$. The associated magnetic pressure, $P_{\mathrm{mag}} = \frac{\mu_0 \chi H^2}{2} \sim \frac{\chi B^2}{2\mu_0}$. For $\chi = 0.05$, $B = 78~\mathrm{mT}$, and $\mu_0 = 4\pi \times 10^{-7}$, we obtain $P_{\mathrm{mag}} \approx 121~\mathrm{Pa}$. The interfacial deformation occurs when magnetic pressure exceed the restoring pressure i.e. $P_{\mathrm{mag}} > \frac{2\sigma}{R} + P_g$. In the present case, $P_{\sigma} + P_g \approx 32.16~\mathrm{Pa}$ at h = 0.7 mm. Since $P_{\mathrm{mag}} > P_{\sigma} + P_g$, hump formation at apex is expected, consistent with experimental observations.
The magnetic Bond number, representing the ratio of magnetic to capillary stresses, is defined as

\begin{equation}
\mathrm{Bo}_m =
\frac{\chi B^2 R}{4 \mu_0 \sigma}.
\label{eq:magnetic_bond}
\end{equation}

The calculated value  of $\mathrm{Bo}_m \approx 4.84$. The droplet aspect ratio at time, t=0, $\frac{h}{R} = \frac{0.7}{1.6} \approx 0.44$, which satisfies $h/R < 1$, corresponding to a flattened sessile droplet. Accounting for finite droplet height, a modified magnetic Bond number is introduced,

\begin{equation}
\mathrm{Bo}_m^{*} = \mathrm{Bo}_m \left(\frac{h}{R}\right),
\label{eq:magnetic_bond2}
\end{equation}

The obtained $\mathrm{Bo}_m^{*} \approx 2.1.$. This confirms noticeable magnetic deformation, validating the experimental results.

The interfacial dynamics of the droplet during the magnet switched-OFF state are governed by the capillary--inertial time scale:

\begin{equation}
    \tau = \sqrt{\frac{\rho_f R^3}{\sigma}}
\end{equation}

The calculated value of $\tau = 0.015 \, \text{s}$ and corresponding natural response frequency, $f_r \sim \frac{1}{\tau} \approx 67 \, \text{Hz}$, which is significantly larger than the maximum actuation frequency employed in the present study ($f_{\max}$). This disparity indicates that resonance effects are not expected.

The capillary wavelength, $\lambda_c = 2\pi l_c = 8.7 \, \text{mm}$. Since $\lambda_c \gg R$, the visualized hump like upward  deformation is not driven by a Rosensweig-type instability\cite{cowley1967interfacial,lange2007linear}.

To characterize the droplet’s dynamic response under a time-dependent magnetic field, we introduce a non-dimensional switching frequency:

\begin{equation}
   S_h = f \tau
\end{equation}

The calculated values of $S_h$ = $0.00024,\; 0.00075,\; 0.0015,\; 0.003,\; 0.0075,\; 0.015,\; 0.075$ for corresponding $f$ =0.016,\; 0.05,\; 0.1,\; 0.2,\; 0.5,\; 1,\; 5 \, \text{Hz}, respectively. Because $S_h \ll 1$ throughout the investigated frequency range, the droplet exhibits a quasi-steady response, with no sustained oscillatory motion. The interface therefore restores smoothly following magnetic field removal.

It is important to note that the contact line remains pinned for approximately $80\%$ of the evaporation time. Beyond this stage, the contact line undergoes intermittent shrinkage to a new pinning site, followed by spreading to another site when the magnetic field is switched OFF. This avanche like motion can be attributed to magneto-capillary interactions rather than classical resonance dynamics.

\subsubsection{Mechanism for the distinct deposition morphology}

The deposition morphology of an evaporating colloidal sessile droplet is governed by particle transport and contact line dynamics. The spatial distribution of the magnetic field inside the droplet plays a critical role in regulating particle motion.

\begin{figure*}[htb!]
    \centering
    \includegraphics[width=1\linewidth]{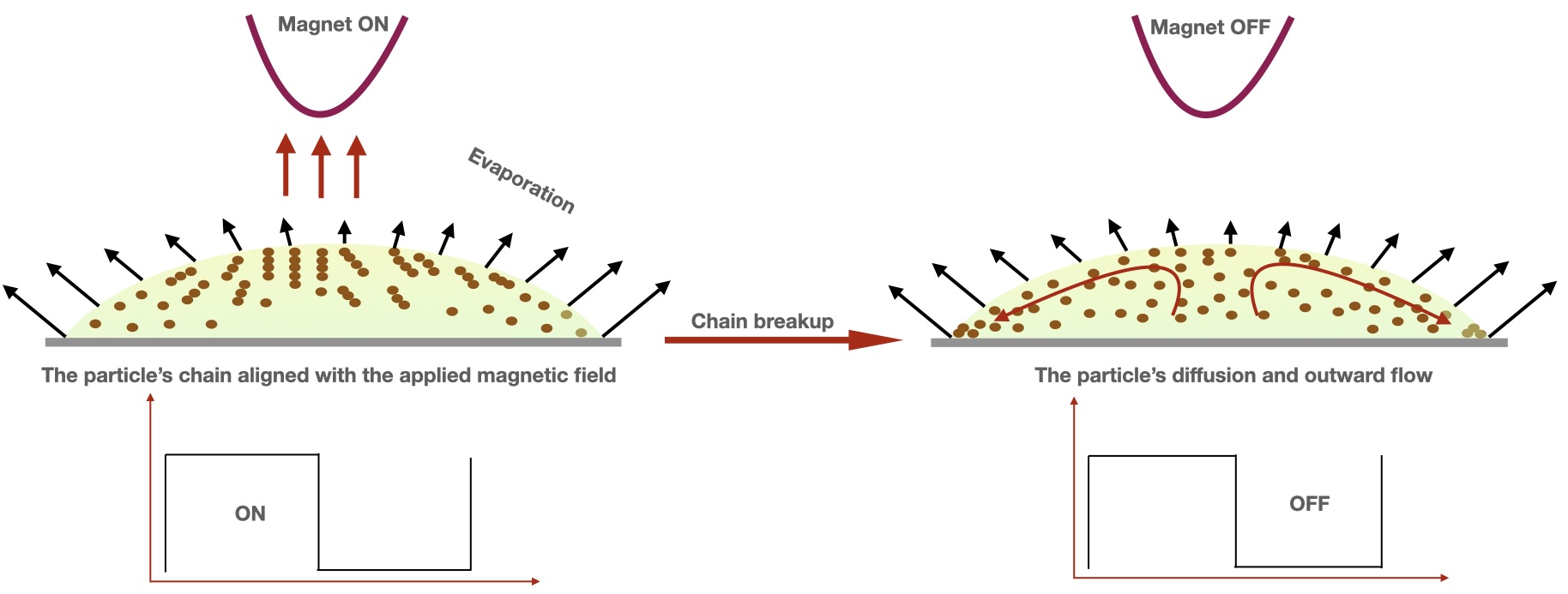}
    \caption{Illustration of the chain formation and its breakup corresponding to magnet ON and magnet OFF, respectively.}
    \label{fig:chain}
\end{figure*}

The magnetic field strength is weaker near the contact line (Figure \ref{fig:Height}); therefore, the magnetic forces acting on the particles are insufficient to counterbalance the capillary-driven outward flow. As a result, particles are advected toward the contact line and progressively deposited during the pinned evaporation stage up to $t/t_0 \approx 0.8$ \cite{deegan1997capillary,deegan2000contact}.

Figure \ref{fig:chain} illustrates particle behavior during the magnet switch-ON and switch-OFF phases. When the magnetic field is applied, particles align along the field direction, leading to chain formation. During the switch-OFF duration, these chains break, enabling particle diffusion within the droplet. In this state, particle transport is governed by evaporation-induced interfacial flows.

The particle motion during the switch-OFF duration is influenced by capillary flow and Marangoni convection, as previously demonstrated \cite{saroj2021magnetophoretic}. The flow reversal driven by Marangoni stresses diminishes as the droplet becomes thin ($h/R \ll 1$), beyond which capillary flow dominates. This results in outward radial transport and particle accumulation at the peripheral region. Notably, Marangoni convection is significantly suppressed under magnetic field application \cite{saroj2021magnetophoretic}. Consequently, particle diffusion becomes the dominant redistribution mechanism during the magnetic field removal phase.

Figure \ref{fig:dia} shows that the contact line exhibits avalanche-like motion induced by magnetocapillary effects after approximately 80\% of the total evaporation time. During the magnetic field switch-ON state, the contact line undergoes rapid contraction, whereas during the switch-OFF state it spreads to establish a new pinning state. The corresponding contact line diameters are denoted as $d_{\mathrm{ON}}$ and $d_{\mathrm{OFF}}$, respectively.

Both $d_{\mathrm{ON}}$ and $d_{\mathrm{OFF}}$ decrease monotonically with $t/t_0$, as shown in Figure \ref{fig:dia}(a), with $d_{\mathrm{ON}}$ consistently exceeding $d_{\mathrm{OFF}}$. During the switch-OFF state ($d_{\mathrm{OFF}}$), particles migrate toward the contact line, promoting ring formation. The spacing and thickness of the rings are strongly dependent on the magnetic switching frequency.

The diffusion time scale and magnetic field distribution collectively regulate particle deposition behavior, including ring density and spacing. Particle diffusion during the magnetic field removal phase is characterized by

\begin{equation}
\tau_D = \frac{\rho_f R^2}{\mu_f}.
\end{equation}

For $R = 1.6 \, \mathrm{mm}$, the diffusion time scale is estimated as $\tau_D \approx 2.93 \, \mathrm{s}$. The competition between magnetic actuation and particle diffusion can be quantified using the magnetic switching number,

\begin{equation}
S_D = f \tau_D.
\end{equation}

Complete particle diffusion during the switch-OFF state occurs when $S_D \leq 1$.

Figure \ref{fig:fn} shows the variation of the number of concentric rings ($N_r$) as a function of $S_D$. The ring number increases with $S_D$ up to a critical value $S_{Dc} = 0.586$, corresponding to an optimum actuation frequency $f_c = 0.2 \, \mathrm{Hz}$. Up to this condition, the ring number follows the scaling

\begin{equation}
N_r \sim 14.125 \, S_D^{0.66},
\end{equation}

with $R^2 = 0.997$, indicating excellent agreement with experimental observations.

\begin{figure}
    \centering
    \includegraphics[width=1\linewidth]{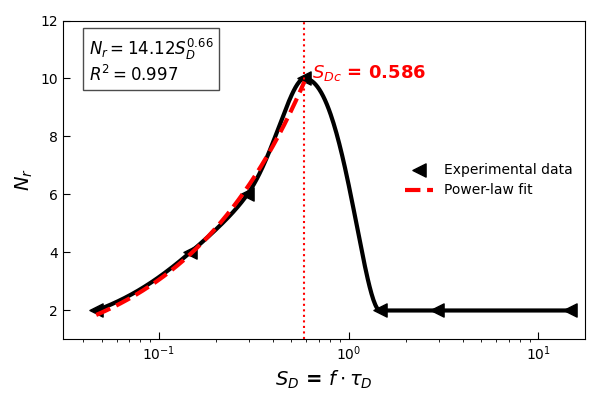}
    \caption{Number of concentric ring ($N_r$) as a function of non--dimensional switching number $S_D$.}
    \label{fig:fn}
\end{figure}

Beyond the optimum frequency, the number of rings decreases due to insufficient diffusion time during the switch-OFF state. Particles are therefore unable to migrate effectively toward the contact line, weakening pinning reinforcement. Instead, particles are redistributed by contact line motion upon magnetic field reactivation.

As evaporation progresses, the local particle concentration within the droplet increases. For $f > f_c$, incomplete chain breakup and restricted diffusion lead to suppression of multi-ring formation, resulting primarily in thick peripheral deposition and enhanced central accumulation. The dense central deposition observed during the final evaporation stage arises from increased local particle concentration and attractive particle--substrate interactions \cite{saroj2019drying}.

\begin{figure}
    \centering
    \includegraphics[width=1\linewidth]{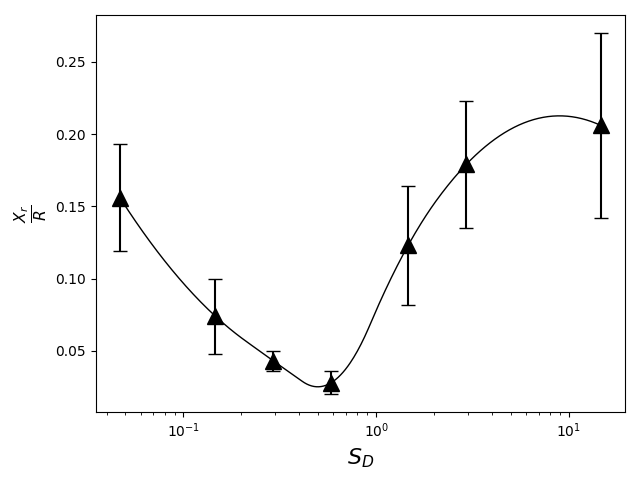}
    \caption{The spacing between the concentric ring ($X_r$) normalized by 'R' as a function of magnetic switching number $S_D$.}
    \label{fig:r}
\end{figure}

Figure \ref{fig:r} shows the spacing between concentric rings ($X_r$) as a function of $S_D$. The spacing is measured from the outer edge of one ring to the inner edge of the subsequent ring. For each frequency, the spacing values are averaged, and the standard deviation is represented as error bars. The ring spacing decreases with increasing frequency up to $f_c$, beyond which it increases.

Using the averaged spacing, the drift velocity is defined as

\begin{equation}
U_d = f X_r.
\end{equation}

The transition in drift velocity scaling at $f_c$ reflects a shift in the dominant particle transport mechanism. For $f < f_c$, the weak scaling $U_d \sim f^{0.31}$ indicates diffusion-dominated particle redistribution. In this regime, the magnetic actuation timescale remains comparable to or longer than the diffusion timescale, enabling particle relaxation within each cycle.

particle.
\begin{figure}
    \centering
    \includegraphics[width=1\linewidth]{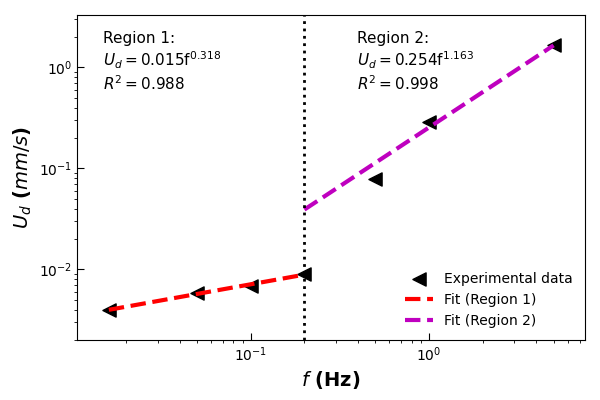}
    \caption{Drift velocity ($U_d$) vs $f$}
    \label{fig:fn}
\end{figure}

For $f > f_c$, the scaling steepens to $U_d \sim f^{1.163}$, indicating dynamically constrained particle motion. Here, the actuation timescale becomes shorter than the diffusion timescale, restricting effective particle redistribution. Consequently, coherent radial migration is suppressed, leading to multi-ring attenuation.

The scaling law predicts $N_r \sim f^{0.66}$. Since $N_r \sim \frac{R}{X_r}, \quad \text{and} \quad X_r \sim \frac{U_d}{f}$, we obtain $N_r \sim \frac{R f}{U_d}$, by substituting the experimentally observed scaling $U_d \sim f^{0.32}$ yields $N_r \sim f^{0.68}$, which is consistent with experimental measurements.

These results demonstrate that the magnetic switching number provides a robust framework for interpreting frequency-mediated transitions in deposition morphology.


\section{Conclusion}

The present study establishes a systematic approach to regulate deposition morphology in an evaporating ferrofluid droplet on a glass substrate using a time-dependent magnetic field. Interfacial deformation accompanied by hump formation at the droplet apex is observed during magnetic field application. The underlying deformation mechanism is presented and validated through experimental observations.

A quasi-steady droplet response is identified and supported by calculations of the capillary time scale and the non-dimensional magnetic actuation time following removal of the magnetic field. The contact line remains pinned for approximately 80\% of the total evaporation time. Subsequently, the contact line exhibits avalanche-like motion, contracting during the switched ON state and spreading during the switched OFF state.

Multi-concentric ring patterns form due to diffusive particle transport during the switched OFF state. The diffusion time scale is evaluated, and an associated non-dimensional magnetic switching number, $S_D$, is proposed. This magnetic switching number serves as a governing control parameter for multi-ring formation. The number of rings initially increases with actuation frequency up to $f_c = 0.2$ Hz. Beyond this critical frequency, the ring count decreases, and coffee-ring suppression is promoted, resulting in more uniform central deposition.

Overall, the conventional coffee-ring morphology is transformed into multi-concentric ring patterns under time-dependent magnetic actuation, followed by suppression at higher actuation frequencies.


\bibliography{reference}

@PREAMBLE{
 "\providecommand{\noopsort}[1]{}" 
 # "\providecommand{\singleletter}[1]{#1}%" 
}

@article{seifert2015additive,
  title={Additive manufacturing technologies compared: morphology of deposits of silver ink using inkjet and aerosol jet printing},
  author={Seifert, Tobias and Sowade, Enrico and Roscher, Frank and Wiemer, Maik and Gessner, Thomas and Baumann, Reinhard R},
  journal={Industrial \& Engineering Chemistry Research},
  volume={54},
  number={2},
  pages={769--779},
  year={2015},
  publisher={ACS Publications}
}

@article{lim2009experimental,
  title={Experimental study on spreading and evaporation of inkjet printed pico-liter droplet on a heated substrate},
  author={Lim, Taewoong and Han, Sewoon and Chung, Jaewon and Chung, Jin Taek and Ko, Seunghwan and Grigoropoulos, Costas P},
  journal={International Journal of Heat and Mass Transfer},
  volume={52},
  number={1-2},
  pages={431--441},
  year={2009},
  publisher={Elsevier}
}

@article{he2017controlling,
  title={Controlling coffee ring formation during drying of inkjet printed 2D inks},
  author={He, Pei and Derby, Brian},
  journal={Advanced Materials Interfaces},
  volume={4},
  number={22},
  pages={1700944},
  year={2017},
  publisher={Wiley Online Library}
}

@article{kolchanov2019sol,
  title={Sol--gel magnetite inks for inkjet printing},
  author={Kolchanov, Denis S and Slabov, Vladislav and Keller, Kirill and Sergeeva, Ekaterina and Zhukov, Mikhail V and Drozdov, Andrey S and Vinogradov, Alexandr V},
  journal={Journal of Materials Chemistry C},
  volume={7},
  number={21},
  pages={6426--6432},
  year={2019},
  publisher={Royal Society of Chemistry}
}

@article{al2019inkjet,
  title={Inkjet printing of magnetic particles toward anisotropic magnetic properties},
  author={Al-Milaji, Karam Nashwan and Hadimani, Ravi L and Gupta, Shalabh and Pecharsky, Vitalij K and Zhao, Hong},
  journal={Scientific Reports},
  volume={9},
  number={1},
  pages={16261},
  year={2019},
  publisher={Nature Publishing Group UK London}
}

@article{saroj2020magnetic,
  title={Magnetic suppression of the coffee ring effect},
  author={Saroj, Sunil Kumar and Panigrahi, Pradipta Kumar},
  journal={Journal of Magnetism and Magnetic Materials},
  volume={513},
  pages={167199},
  year={2020},
  publisher={Elsevier}
}

@article{song2014inkjet,
  title={Inkjet printing of magnetic materials with aligned anisotropy},
  author={Song, Han and Spencer, Jeremy and Jander, Albrecht and Nielsen, Jeffrey and Stasiak, James and Kasperchik, Vladek and Dhagat, Pallavi},
  journal={Journal of Applied Physics},
  volume={115},
  number={17},
  year={2014},
  publisher={AIP Publishing}
}

@article{wu2018multi,
  title={Multi-ring deposition pattern of drying droplets},
  author={Wu, Mengmeng and Man, Xingkun and Doi, Masao},
  journal={Langmuir},
  volume={34},
  number={32},
  pages={9572--9578},
  year={2018},
  publisher={ACS Publications}
}

@article{moffat2009effect,
  title={Effect of TiO2 nanoparticles on contact line stick- slip behavior of volatile drops},
  author={Moffat, J Ross and Sefiane, Khellil and Shanahan, Martin ER},
  journal={The Journal of Physical Chemistry B},
  volume={113},
  number={26},
  pages={8860--8866},
  year={2009},
  publisher={ACS Publications}
}

@article{saroj2021magnetophoretic,
  title={Magnetophoretic control of diamagnetic particles inside an evaporating droplet},
  author={Saroj, Sunil Kumar and Panigrahi, Pradipta Kumar},
  journal={Langmuir},
  volume={37},
  number={51},
  pages={14950--14967},
  year={2021},
  publisher={ACS Publications}
}

@article{weon2010capillary,
  title={Capillary force repels coffee-ring effect},
  author={Weon, Byung Mook and Je, Jung Ho},
  journal={Physical Review E—Statistical, Nonlinear, and Soft Matter Physics},
  volume={82},
  number={1},
  pages={015305},
  year={2010},
  publisher={APS}
}

@article{saroj2016two,
  title={Two-fluid mixing inside a sessile micro droplet using magnetic beads actuation},
  author={Saroj, Sunil Kumar and Asfer, Mohammed and Sunderka, Aman and Panigrahi, Pradipta Kumar},
  journal={Sensors and Actuators A: Physical},
  volume={244},
  pages={112--120},
  year={2016},
  publisher={Elsevier}
}

@article{deegan1997capillary,
  title={Capillary flow as the cause of ring stains from dried liquid drops},
  author={Deegan, Robert D and Bakajin, Olgica and Dupont, Todd F and Huber, Greb and Nagel, Sidney R and Witten, Thomas A},
  journal={Nature},
  volume={389},
  number={6653},
  pages={827--829},
  year={1997},
  publisher={Nature Publishing Group UK London}
}

@article{deegan2000contact,
  title={Contact line deposits in an evaporating drop},
  author={Deegan, Robert D and Bakajin, Olgica and Dupont, Todd F and Huber, Greg and Nagel, Sidney R and Witten, Thomas A},
  journal={Physical review E},
  volume={62},
  number={1},
  pages={756},
  year={2000},
  publisher={APS}
}

@article{saroj2019drying,
  title={Drying pattern and evaporation dynamics of sessile ferrofluid droplet on a PDMS substrate},
  author={Saroj, Sunil Kumar and Panigrahi, Pradipta Kumar},
  journal={Colloids and Surfaces A: Physicochemical and Engineering Aspects},
  volume={580},
  pages={123672},
  year={2019},
  publisher={Elsevier}
}

@book{bashtovoi1987magnetic,
  title={Magnetic Fluids and Applications Handbook},
  author={Bashtovoi, V. G. and Berkovsky, B. M. and Vislovich, A. N.},
  year={1987},
  publisher={Hemisphere}
}

@article{dormann1997magnetic,
  title={Magnetic relaxation in fine-particle systems},
  author={Dormann, J. L. and Fiorani, D. and Tronc, E.},
  journal={Advances in Chemical Physics},
  volume={98},
  pages={283--494},
  year={1997}
}

@article{marc2022coffee,
  title={Coffee-ring formation through the use of the multi-ring mechanism guided by the self-assembly of magnetic nanoparticles},
  author={Mar{\'c}, M and Wolak, W and Drzewi{\'n}ski, A and Dudek, MR},
  journal={Scientific Reports},
  volume={12},
  number={1},
  pages={20131},
  year={2022},
  publisher={Nature Publishing Group UK London}
}

@incollection{kumar2017fabrication,
  title={Fabrication of nanostructures with bottom-up approach and their utility in diagnostics, therapeutics, and others},
  author={Kumar, Sanjay and Bhushan, Pulak and Bhattacharya, Shantanu},
  booktitle={Environmental, chemical and medical sensors},
  pages={167--198},
  year={2017},
  publisher={Springer}
}

@article{mampallil2015acoustic,
  title={Acoustic suppression of the coffee-ring effect},
  author={Mampallil, Dileep and Reboud, Julien and Wilson, Rab and Wylie, Douglas and Klug, David R and Cooper, Jonathan M},
  journal={Soft matter},
  volume={11},
  number={36},
  pages={7207--7213},
  year={2015},
  publisher={Royal Society of Chemistry}
}

@article{yang2020air,
  title={Air bubble-triggered suppression of the coffee-ring effect},
  author={Yang, Quansan and Lv, Cunjing and Hao, Pengfei and He, Feng and Ouyang, Yuanyuan and Niu, Fenglei},
  journal={Colloid and Interface Science Communications},
  volume={37},
  pages={100284},
  year={2020},
  publisher={Elsevier}
}

@article{sliz2020taming,
  title={Taming the coffee ring effect: Enhanced thermal control as a method for thin-film nanopatterning},
  author={Sliz, Rafal and Czajkowski, Jakub and Fabritius, Tapio},
  journal={Langmuir},
  volume={36},
  number={32},
  pages={9562--9570},
  year={2020},
  publisher={ACS Publications}
}

@article{eral2011suppressing,
  title={Suppressing the coffee stain effect: how to control colloidal self-assembly in evaporating drops using electrowetting},
  author={Eral, H Burak and Augustine, D Mampallil and Duits, Michael HG and Mugele, Frieder},
  journal={Soft Matter},
  volume={7},
  number={10},
  pages={4954--4958},
  year={2011},
  publisher={Royal Society of Chemistry}
}

@article{lange2007linear,
  title={Linear and nonlinear approach to the Rosensweig instability},
  author={Lange, Adrian and Richter, Reinhard and Tobiska, Lutz},
  journal={GAMM-Mitteilungen},
  volume={30},
  number={1},
  pages={171--184},
  year={2007},
  publisher={Wiley Online Library}
}

@article{cowley1967interfacial,
  title={The interfacial stability of a ferromagnetic fluid},
  author={Cowley, MD and Rosensweig, Ronald E},
  journal={Journal of Fluid mechanics},
  volume={30},
  number={4},
  pages={671--688},
  year={1967},
  publisher={Cambridge University Press}
}

@article{men2015controlled,
  title={Controlled evaporative self-assembly of Fe 3 O 4 nanoparticles assisted by an external magnetic field},
  author={Men, Yonghong and Wang, Wenqin and Xiao, Peng and Gu, Jincui and Sun, Aihua and Huang, Youju and Zhang, Jiawei and Chen, Tao},
  journal={RSC Advances},
  volume={5},
  number={40},
  pages={31519--31524},
  year={2015},
  publisher={Royal Society of Chemistry}
}

@article{erb2013self,
  title={Self-shaping composites with programmable bioinspired microstructures},
  author={Erb, Randall M and Sander, Jonathan S and Grisch, Roman and Studart, Andr{\'e} R},
  journal={Nature communications},
  volume={4},
  number={1},
  pages={1712},
  year={2013},
  publisher={Nature Publishing Group UK London}
}

@article{yang2023recent,
  title={Recent advances in magnetically responsive photonic crystals assembled by anisotropic building blocks: Synthesis, challenges and outstanding applications},
  author={Yang, Shuying and Ding, Rongmin and Ma, Ranran and Wu, Mengyi and Chen, Pei and Zhang, Yajie and Ye, Aoli and You, Linjun and Xiao, Deli},
  journal={Journal of Magnetism and Magnetic Materials},
  volume={585},
  pages={171097},
  year={2023},
  publisher={Elsevier}
}

@book{maier2007plasmonics,
  title={Plasmonics: fundamentals and applications},
  author={Maier, Stefan A and others},
  volume={1},
  year={2007},
  publisher={Springer}
}

@article{tiberto2013magnetic,
  title={Magnetic properties of jet-printer inks containing dispersed magnetite nanoparticles},
  author={Tiberto, Paola and Barrera, Gabriele and Celegato, Federica and Co{\"\i}sson, Marco and Chiolerio, Alessandro and Martino, Paola and Pandolfi, Paolo and Allia, Paolo},
  journal={The European Physical Journal B},
  volume={86},
  number={4},
  pages={173},
  year={2013},
  publisher={Springer}
}

@article{bhattacharjee2026evaporative,
  title={Evaporative self-assembly of nanocolloids: fundamentals, patterns and applications},
  author={Bhattacharjee, Suman and Vaisakh, NP and Khawas, Sanjoy and Afnan, Urba and Srivastava, Sunita},
  journal={Journal of Physics: Condensed Matter},
  volume={38},
  number={6},
  pages={063001},
  year={2026},
  publisher={IOP Publishing}
}

@article{pinto2020application,
  title={Application of magnetic nanoparticles for water purification},
  author={Pinto, Mariana and Ramalho, PSF and Moreira, NFF and Gon{\c{c}}alves, AG and Nunes, OC and Pereira, MFR and Soares, OSGP},
  journal={Environmental Advances},
  volume={2},
  pages={100010},
  year={2020},
  publisher={Elsevier}
}

@article{yellen2005arranging,
  title={Arranging matter by magnetic nanoparticle assemblers},
  author={Yellen, Benjamin B and Hovorka, Ondrej and Friedman, Gary},
  journal={Proceedings of the national academy of sciences},
  volume={102},
  number={25},
  pages={8860--8864},
  year={2005},
  publisher={National Academy of Sciences}
}

\end{document}